\documentclass[amsmath, amsthm, amssymb, floatfix, reprint, twocolumn, notitlepage, nofootinbib, 10pt]{revtex4-1}% this article uses the Revtex 4-1 format
\usepackage{setspace}
\usepackage{amsmath}
\usepackage{breqn}
\usepackage{graphicx}
\usepackage[nearskip,margin = 0pt,caption=false]{subfig}

\usepackage{verbatim}
\usepackage{amsfonts}
\usepackage{amssymb}
\usepackage{epstopdf}
\usepackage{lipsum}
\usepackage{siunitx}
\usepackage{xcolor}
\usepackage{array}
\usepackage{braket}
\usepackage[normalem]{ulem}
\usepackage[resetlabels,labeled]{multibib}
\DeclareGraphicsExtensions{.pdf,.eps,.png,.jpg,.mps}
\usepackage{etoolbox}
\usepackage{subfiles} 
\newcommand{\norm}[1]{\left\Vert{#1}\right\Vert}

\newcommand{\hl}[1]{{\color{black} #1}}

\begin{document}
\title{Superradiance in dynamically modulated Tavis-Cummings model with spectral disorder}
\author{A.D. White$^{1,\dagger}$, R. Trivedi$^{2,3,\dagger}$, K. Narayanan$^{1}$, J.Vu\v{c}kovi\'{c}$^{1}$\\
\vspace{+0.05 in}
$^1$E. L. Ginzton Laboratory, Stanford University, Stanford, CA, USA.\\
$^2$Max Planck Institute of Quantum Optics, M\"unchen, Germany.\\
$^3$Department of Electrical and Computer Engineering, University of Washington, Seattle, USA \\
{\small $^{\dagger}$ Equal contribution; adwhite@stanford.edu, rtriv@uw.edu}}

\begin{abstract}
    \noindent
    Superradiance is the enhanced emission of photons from quantum emitters collectively coupling to the same optical mode. However, disorder in the resonant frequencies of the quantum emitters can perturb this effect. In this paper, we study the interplay between superradiance and spectral disorder in a dynamically modulated Tavis-Cummings model. Through numerical simulations and analytical calculations, we show that the effective cooperativity of the superradiant mode, which is always formed over an extensive number of emitters, can be multiplicatively enhanced with a quantum control protocol modulating the resonant frequency of the optical mode. Our results are relevant to experimental demonstration of superradiant effects in solid-state quantum optical systems, wherein the spectral disorder is a significant technological impediment towards achieving photon-mediated emitter-emitter couplings. 
    % Nonclassical states of light are an essential resource for quantum communication and metrology, and the development of higher fidelity and higher entanglement photon sources is critical to advance the state of the art in both fields. Solid state emitters are ideal candidates for such non-classical sources, as they are excellent sources of single photons, can be integrated in solid state devices, and do not need to be cooled to milli-kelvin temperatures. However, the natural inhomogeneous broadening and low coopertivity of state emitters make coupling emitters to generate complex photon states such as dicke superradiant states extraordinarily difficult in practice.
    % Here we propose a quantum control technique that uses frequency modulation to restore the cooperative enhancement of ensembles of emitters in the presence of inhomogeneity.
    % Through adjoint optimization of a drive applied to a cavity QED system, we are able to enhance the overlap of the output photon state of a collection of inhomogenous emitters with that of an ideal Dicke superradiant state by several orders of magnitude. Additionally, we provide a heuristic optimization that scales polynomially with the number of emitters and can still boost the superradiant overlap by two orders of magnitude.
\end{abstract}

  \maketitle
Coherent interaction between a collection of quantum emitters and an optical mode, theoretically described by the Tavis-Cummings model \cite{tavis1968exact}, has been a topic of intense theoretical interest since the conception of quantum optics. Multiple emitters coupling strongly to the same optical mode are known to exhibit cooperative effects, the most prominent of which is the formation of a superradiant state \cite{ gross1982superradiance, rehler1971superradiance, scully2009super, dicke1954coherence}. Superradiant states are fully symmetric collective excitations of the multiple emitters whose interaction with the optical mode is enhanced due to a constructive interference of the individually emitted photons. These states underlie the physical phenomena of collective spontaneous emission (Dicke superradiance) \cite{dicke1954coherence, andreev1980collective, clemens2003collective, eleuch2014open} and superradiant phase transitions \cite{hepp1973superradiant, lambert2004entanglement, bamba2016superradiant}. Furthermore, the enhancement of light-matter interaction in such systems has implications for design and implementation of a number of quantum information processing blocks, such as transducers \cite{casabone2015enhanced}, memories \cite{ortiz2018experimental} and non-classical light sources \cite{jahnke2016giant}.

However, technologically relevant experimental systems that can potentially demonstrate and use superradiance, such as solid-state quantum optical systems like quantum dots \cite{scheibner2007superradiance, grim2019scalable}, color centers \cite{angerer2018superradiant} and rare-earth ions \cite{zhong2017interfacing}, often suffer from spectral disorder amongst the quantum emitters. 
Since spectral disorder can disrupt interactions between emitters, it competes with the collective emitter-optical mode interaction and can prevent the formation of the superradiant state.
%Since this spectral disorder can disrupt the collective interaction between the different quantum emitters, it competes with the collective emitter-optical mode interaction emitter in the formation of the superradiant state. 
This interplay between disorder and coherent interaction has been extensively studied in time-independent quantum systems arising in many-body physics \cite{anderson1958absence, lee1981anderson, alet2018many, nandkishore2015many, pal2010many, nandkishore2017many, nag2019many, smith2016many} as well as quantum optics \cite{moreira2019localization, akkermans2008photon, ashhab2017superradiance, biella2013subradiant, cottier2018role, kelly2021effect, manzoni2017simulating, fayard2021many, temnov2005superradiance, trivedi2019photon, mivehvar2017disorder, debnath2019collective} 
There has been recent interest in understanding the impact of dynamical modulation, applied globally on all the emitters, on the properties of such models. Such modulation schemes are easily experimentally accessible, e.g. in quantum optical systems with a dynamically modulated optical mode (such as modulation of the resonant frequency of a cavity \cite{zhang2019electronically}) or with collectively modulated emitters (through simultaneously laser driving or stark shifting the resonant frequencies of all emitters \cite{sun2018cavity, lukin2020spectrally}). These global quantum controls designed with off-the-shelf optimization techniques \cite{de2011second, d2007introduction} have been be used to potentially compensate for, or in some cases exploit, the spectral disorder \cite{julsgaard2013quantum, gorshkov2008photon} for building quantum information processing hardware such as quantum transducers and memories. However, an understanding of their effectiveness in compensating disorder, specially in the thermodynamic limit of large number of emitters, is less well understood.
%and retardation effects \cite{sinha2020non, masson2020many} on collective phenomena such as superradiance. 

%As superradiance can be leveraged by technologically relevant systems, 
%Because of its deleterious effects on superradiance formation, it is 
%In light of 

In this paper, we study the interplay of superradiance and spectral disorder in a dynamically modulated Tavis-Cummings model. Within the single-photon subspace, the all-to-all coupling between the emitters enables the formation of a superradiant state over an extensive number of emitters irrespective of the extent of disorder in the system. We study the impact of the dynamical modulation, designed as a quantum control to compensate for the spectral disorder in the system, on the formation of this superradiant state. We demonstrate that this dynamical modulation can achieve a multiplicative enhancement in the cooperativity of the superradiant state even in the limit of a large number of emitters. 
%We provide a theoretical explanation of this phenomenon by analytically constructing a control pulse that achieves this improvement in the thermodynamic limit. Furthermore, by theoretically providing an upper bound on the expected improvement, we argue that this control pulse achieves near optimal scaling of the improvement with the extent of spectral disorder. 
Finally, we provide evidence that this control pulse, designed by only considering single-photon dynamics, also multiplicatively enhances superradiance in the multi-photon subspaces of the Tavis-Cummings model.

We consider $N$ emitters coupled to a cavity with the following Hamiltonian (Fig 1a)
\begin{align}
    &H = \omega_c(t) a^\dagger a + H_e + H_c, \ \text{where} \nonumber\\
    &H_e = \sum_{i = 1}^{N}\omega_i\sigma_i^\dagger\sigma_i \text{ and } H_c =  g(N)\sum_{i = 1}^{N}(a^\dagger\sigma_i + \sigma_i^\dagger a)
\end{align}
where $\omega_c(t)$ is the time dependent frequency of the cavity, $\omega_i$ is the resonant frequency of the $i$th emitter, $a^\dagger$ is the raising operator of the cavity, $\sigma_i^\dagger$ is the excitation operator of the $i$th emitter, and $g(N)$ is the emitter-cavity coupling strength (which we allow to depend on the number of emitters). Furthermore, we assume that the cavity emits into an output channel with decay rate $\kappa$, and the emitters in addition to coupling to the cavity also individually decay with decay rate $\gamma$. We will denote the ground state of this model, with no photons in the cavity and all emitters in their individual ground state, by $\ket{\text{G}}$.

% To quantify the superradiant behavior of the system, we use two metrics, eigenstate superradiance and single photon generation fidelity. We define eigenstate superradiance as the maximum overlap of the system eigenstates with the superradiant eigenstate of the homogenous system,  $\max_i \left|\braket{\phi_{sr} | v_i}\right|^2$.
% We define the fidelity as the probability that the system, initialized in an equal superposition of all emitter excited states, emits a single photon in the presence of an additional nonradiative decay, $\int_0^\infty |\psi_{out}(t)|^2 dt$.

To quantify the superradiant behavior of this system, we use two metrics: \emph{eigenstate superradiance} and \emph{photon generation fidelity}. The eigenstate superradiance ($\text{ES}$) is calculated as
\begin{align}
    \mu_{\text{ES}}[\omega_c(t)] = \max_{\ket{\phi}} \bigg|\bra{\text{G}}\sum_{i=1}^N \sigma_i \ket{\phi}\bigg|^2,
\end{align}
where the maximization is done over all $\ket{\phi}$, the single-photon Floquet eigenstates of $H(t)$. This metric can be interpreted as a measure of the coupling between the (most) superradiant state and the ground state $\ket{\text{G}}$ induced by $H_c$. We point out that in the absence of disorder and with the emitters being on-resonance with the cavity mode, $\mu_{\text{ES}}= N / 2$. Additionally, we calculate the photon generation fidelity; we initialize the emitters to an initial symmetric state (i.e. $\sum_{i = 1}^N \sigma_i^\dagger \ket{\text{G}} / \sqrt{N}$), allow it to decay into the output channel through the cavity, and compute the probability of a photon being emitted into the output channel:
\begin{align}\label{eq:sph_fid}
    \mu_{\text{FID}}[\omega_c(t)] = \kappa \int_0^\infty \langle a^\dagger(t) a(t)\rangle dt.
\end{align}

We begin by studying the behavior of an unmodulated system ($\omega_c(t) = 0$) within the single photon subspace as a function of $N$. 
\hl{Fig.~\ref{fig:Fig1}b considers a setting where the coupling strength between the cavity and emitters is independent of $N$ (i.e. $g(N) \sim$ constant) -- in this case within the single-photon subspace, $\norm{H_c} \sim g(N)\sqrt{N} \sim \sqrt{N}$ and $\norm{H_e} \sim \text{constant}$. Hence, for large $N$ the dynamics of this system is completely dominated by $H_c$ and spectral disorder does not play any role. This is evidenced both by the eigenvalue superradiance approaching $N/2$ as $N\to \infty$ and the single-photon generation fidelity approaching the fidelity of the homogeneous system (Fig.~\ref{fig:Fig1}b). A more interesting setting, studied in Fig.~\ref{fig:Fig1}c, is where $g(N) \sim 1 / \sqrt{N}$ so as to make the norms $\norm{H_c}$ and $\norm{H_e}$ comparable at large $N$.}
%($g(N)\sqrt{N} = G = $ constant). 
In this case, while superradiance is not completely recovered as $N\to \infty$, we find that the eigenstate superradiance still scales as $N$ --- consequently, a superradiant state is still formed between an extensive number (but not all) of emitters. Likewise, the single photon generation fidelity approaches a non-zero constant as $N\rightarrow \infty$ --- since the coupling constant vanishes in this limit, this is only possible if an extensive number of emitters are cooperatively emitting into the cavity mode. The formation of a superradiant state over an extensive number of emitters stems from the all-to-all coupling between the emitters mediated by the cavity mode, making it fundamentally different from the impact of disorder in models with local interactions, wherein the number of emitters cooperatively interacting with each other would grow at most logarithmically with $N$ \cite{anderson1958absence, aizenman1993localization}.
\begin{figure}[t]
\centering
\includegraphics[width=1\linewidth]{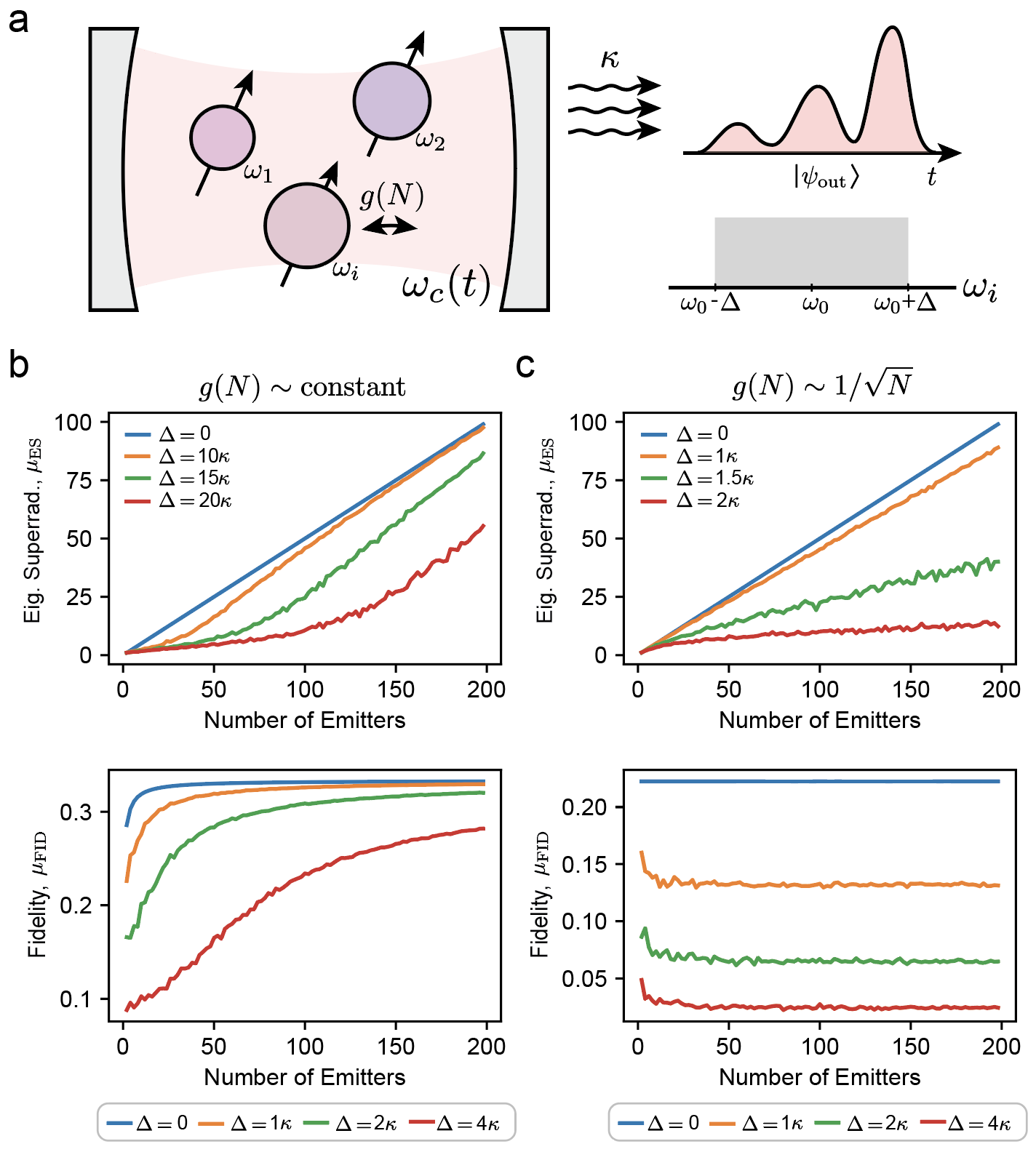}
%\captionsetup{singlelinecheck=no, justification = RaggedRight}
\caption{\label{fig:Fig1}{\bf{Setup and Unmodulated System Scaling}} \hl{\textbf{(a)} We consider $N$ emitters with frequencies $\omega_i$ coupled to a cavity with rate $g(N)$. The cavity is driven by a pulse that imparts a frequency shift $\omega_c(t)$ and is coupled to a waveguide with rate $\kappa$.} \textbf{(b)} Single photon eigenstate superradiance and fidelity of the unmodulated ($\omega_c(t) = 0$) system in (a), with coupling strength $g=\kappa$ held constant. Single photon emission fidelity is calculated in the presence of external decay at rate $2\kappa$. \textbf{(c)} Eigenstate superradiance and fidelity with $g$ scaled as $g = \kappa/\sqrt{N}$. All plots show the average value from 25 different emitter ensembles with frequencies sampled from a uniform distribution.}
\end{figure}

\begin{figure}[t]
\centering
\includegraphics[width=1\linewidth]{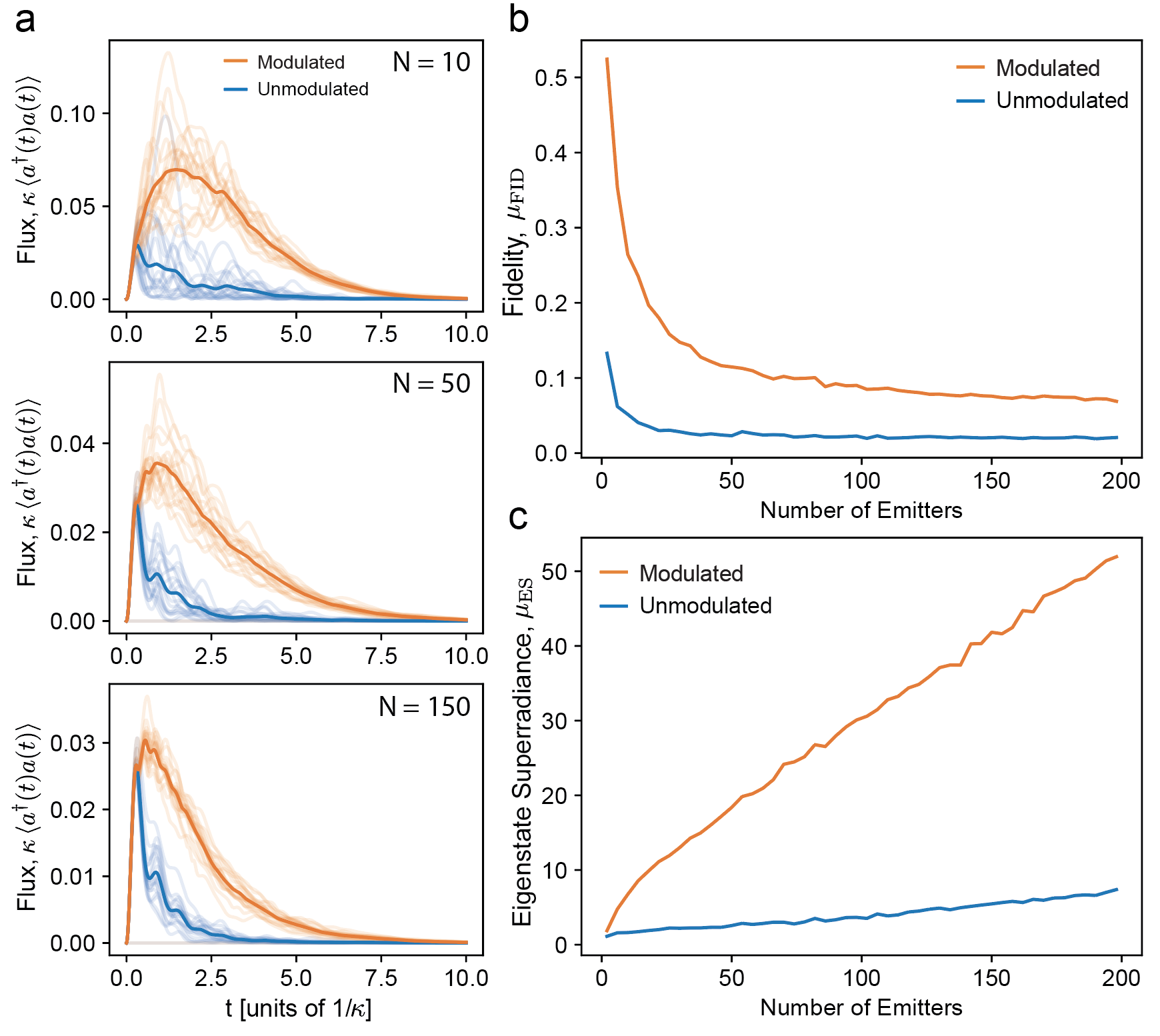}
%\captionsetup{singlelinecheck=no, justification = RaggedRight}
\caption{\label{fig:Fig2}{\hl{\bf{Pulse Optimization}}}
\hl{\textbf{(a)} Output photon flux before and after optimization for $N = 10, 50,$ and $150$ with $g = \kappa/\sqrt{N}$, $\Delta = 10\kappa$, and constant external decay rate $0.5\kappa$. Transparent lines show the photon flux for each ensemble and solid lines show the average. \textbf{(b)} Optimized single photon emission fidelity scaling with $N$ for $g = \kappa/\sqrt{N}$, $\Delta = 10\kappa$, and constant external decay rate $0.5\kappa$. 
%Inset shows ratio of optimized and non-optimized fidelity for 100 emitters across a range of detunings (units of $\kappa$). 
\textbf{(c)} Optimized eigenstate superradiance scaling with $N$ for $g = \kappa/\sqrt{N}$ and $\Delta = 10\kappa$. With these parameters a homogeneous ensemble would exhibit an eigenstate superradiance of $N/2$. 
Plots show the average value from 25 different emitter ensembles.}
%\textbf{(a)} Output single photon wave packet from 50 emitter system with $g = \kappa/\sqrt{50}$ and $\Delta = 8\kappa$ before and after optimization. \textbf{(b)} Optimized eigenstate superradiance scaling with N for $g = \kappa/\sqrt{N}$ and $\Delta = 8\kappa$. With these parameters a homogeneous ensemble would exhibit an eigenstate superradiance of $N/2$. \textbf{(c)} Optimized single photon emission fidelity scaling with N for $g = \kappa/\sqrt{N}$, $\Delta = 8\kappa$, and constant external decay rate $2\kappa$. With these parameters a homogeneous ensemble would exhibit a fidelity of $2/9$. Inset shows ratio of optimized and non-optimized fidelity for 100 emitters across a range of detunings (units of $\kappa$). Plots (b) and (c) show the average value from 25 different emitter ensembles.
}
\end{figure}

We next consider the impact of dynamical modulation of cavity resonance $\omega_c(t)$ on superradiance. %--- in order to calculate a modulation signal $\omega_c(t)$ which best compensates for the spectral disorder, %we choose it to maximize the overlap between the single-photon emitted in the presence of both disorder and modulation and the photon single-photon emitted from a homogeneous system (without disorder or driving). 
\hl{To find a modulation signal that best compensates for the spectral disorder, we maximize the single photon generation fidelity (Eq.~\ref{eq:sph_fid}) with respect to $\omega_c(t)$.
%the overlap between the photon emitted by the modulated disordered system and the photon emitted by an unmodulated homogeneous system (without disorder or driving). 
The optimized modulation signal is computed by using time-dependent scattering theory \cite{trivedi2018few, fischer2018scattering} together with adjoint-sensitivity analysis \cite{cao2003adjoint} (see supplementary for details) to solve the maximization problem. Figure 2a shows the impact that applying this modulation has on the photon emission rate ($\kappa \langle a^\dagger(t) a(t)\rangle$) into the output channel --- we clearly see a significant sustained and consistent enhancement of photon flux due to the modulating signal. }
%As the number of emitters grow, the improvement becomes less variable. 
%we clearly see that the emitted photon wave-packet in the presence of modulation resembles a photon that would be emitted from a homogeneous system when compared with the emission from the unmodulated system. 
Furthermore, as is seen in Figs.~2b and c both the single-photon generation fidelity as well as the eigenstate superradiance (now computed with the eigenstate of the propagator corresponding to the duration for which the pulse is applied) are multiplicatively enhanced when compared to the unmodulated system. 

\begin{figure}[b]
\centering
\includegraphics[width=1\linewidth]{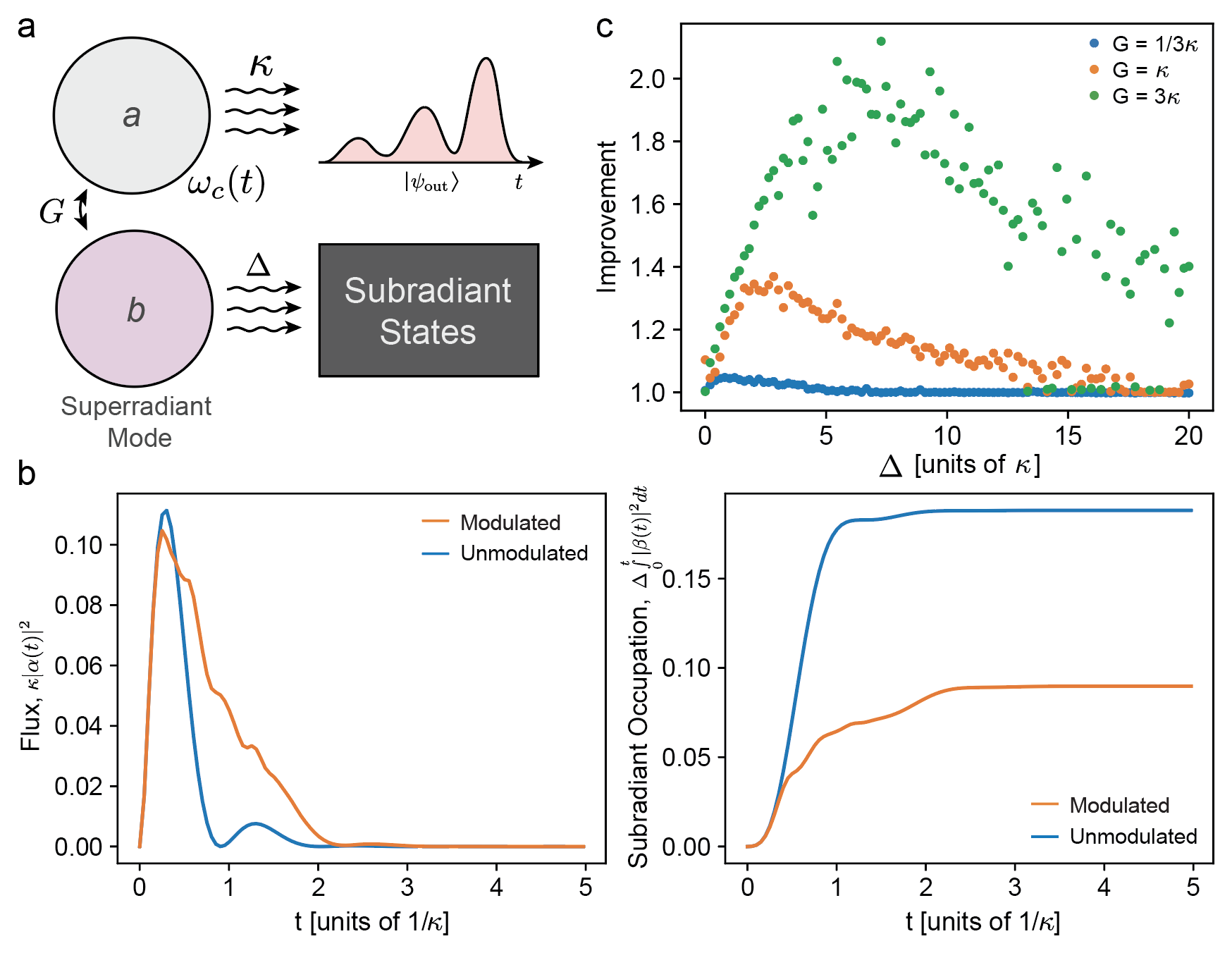}
%\captionsetup{singlelinecheck=no, justification = RaggedRight}
\caption{\label{fig:Fig3}{\hl{\bf{Large $N$ Model}}} \hl{\textbf{(a)} Effective model for the system presented in Fig 1a when emitter frequencies are drawn from a Lorentzian distribution and $N \rightarrow \infty$. 
\textbf{(b)} Output photon flux and subradiant occupation for optimized Large $N$ model with $G = 3\kappa$ and $\Delta = 5\kappa$. \textbf{(c)} Large $N$ optimization improvement scaling with $\Delta$ for $G = 1/3\kappa, 1\kappa,$ and $3\kappa$.}
%\textbf{(b)} Left panel shows photon wavepacket from (a) with $G = 1.5\kappa$ and $\Delta = 4\kappa$ with and without the phase flip pulse applied. Inset shows wavepacket of the homogeneous system. Right panel shows photon wavepacket from a 5000 emitter system with $g = \kappa/\sqrt{5000}$ and $\Delta = 4\kappa$ driven by the same pulse. Inset shows the emitter frequencies obtained by drawing from a truncated lorentzian distribution. \textbf{(c)} Overlap improvement from phase flip pulse scaling with $\Delta$ in the large $N$ model and in large but finite emitter samples. Each point represents the average improvement over 100 emitter ensembles pulled from a truncated Lorentzian distribution. Inset shows value of overlap with phase flip pulse applied over $\Delta$, dashed line shows the fit $0.39/\Delta^2$. }
}
\end{figure}

\begin{figure*}[t]
\centering
\includegraphics[width=1\linewidth]{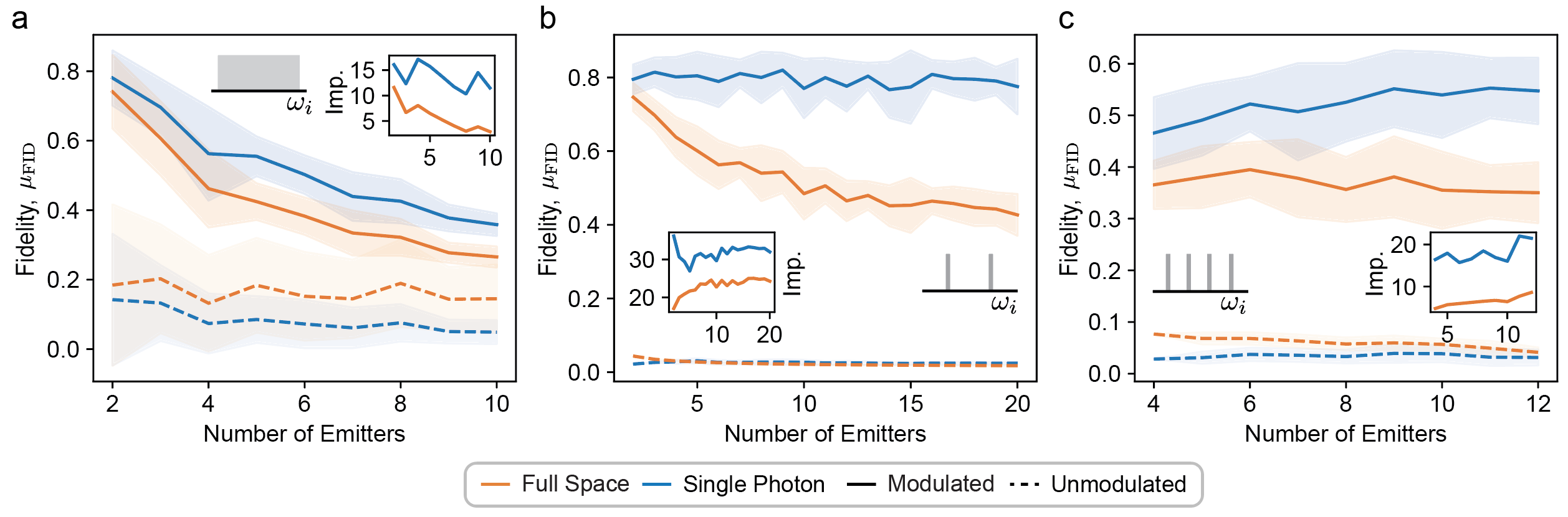}
%\captionsetup{singlelinecheck=no, justification = RaggedRight}
\caption{\label{fig:Fig4}{\hl{\bf{Multi-Photon Optimization}}} \hl{\textbf{(a)} Photon emission fidelity normalized to a homogeneous ensemble for the single photon subspace and full space with emitter frequencies drawn from a uniform distribution. Modulation in the both the single photon and full space are optimized in the single photon subspace. Here $g = \kappa/\sqrt{N}$ and $\Delta = 10\kappa$. Filled area represents the standard deviation across 25 ensembles. Inset shows average improvement by modulating the system with the optimized pulse.
\textbf{(b)} Photon emission fidelity and improvement with emitter frequencies drawn from 2 bins with standard deviation the same as in (a).
\textbf{(c)} Photon emission fidelity and improvement with emitter frequencies drawn from 4 bins with standard deviation the same as in (a).}}
%Decay of a 2 emitter system optimized over the full 2 photon Hilbert space. \textbf{(b)} Decay of a 3 emitter system optimized over the full 3 photon Hilbert space. \textbf{(c)} Schematic of single photon subspace approximation. When N photons are considered, the computational complexity scales with the exponential size of the Hilbert space (top row), whereas the single photon subspace scales only linearly with n (bottom row). \textbf{(d)} Simulations with the single photon subspace approximation with $g = \kappa$. Each data-point represents the geometric mean of 20 optimizations. The emitter frequencies are randomly chosen such that each sample has a mean 0 and standard deviation of $3\kappa$. Plots on the right show histograms of 100 optimizations of 4 emitters.}
\end{figure*}

Surprisingly, a constant enhancement in the superradiance metrics persists even in the limit of a large number of emitters, and improves with an increase in the spectral disorder. To confirm and provide an explanation of this behaviour, we derive an effective analytical model for directly capturing the dynamics in the thermodynamic limit ($N\to \infty$). Assuming a Lorentzian distribution of the emitter frequency [$p(\omega_i) = \Delta_0 / \pi(\omega_i^2 + \Delta_0^2) $], the single-excitation dynamics in the limit of $N\to \infty$ can be captured by a coupled oscillator model (Fig.~3a) ---
\begin{align}
\begin{bmatrix}
    \dot{\alpha}(t) \\
    \dot{\beta}(t)
\end{bmatrix} =
\begin{bmatrix}
    -(i\omega_c(t) + \kappa / 2) & -G \\
    G & -\Delta_0
\end{bmatrix}
\begin{bmatrix}
    \alpha(t) \\
    \beta(t)
\end{bmatrix}
 -iG \begin{bmatrix} e^{-\delta_0 t} \\
  0\end{bmatrix}.
\end{align}
Here $G = \lim_{N\to \infty}g(N) \sqrt{N}$, and $\alpha(t)$ is the amplitude of the dynamically modulated cavity mode, and $\beta(t)$ can be interpreted as the amplitude of the superradiant mode which directly couples to the cavity mode. Note that the inhomogeneous broadening $\Delta_0$ effectively induces a decay in the superradiant mode due to its coupling to the subradiant states due to the inhomogeniety in the emitter frequencies. By optimizing the single photon generation fidelity with this effective model, we find an enhancement in superradiance (Fig 3b) consistent with the finite $N$ results. We clearly see that the application of the modulation reduces the number of photons lost to the subradiant states, and can be viewed as dynamically decoupling the subradiant states from the superradiant mode \cite{viola1999dynamical}. Fig.~3c shows the dependence of the improvement in the photon-generation fidelity achieved by the modulation on the coupling strength $G$ and the broadening $\Delta$ ---  as is intuitively expected, higher improvement in superradiance is seen for more strongly coupled systems. Furthermore, the improvement in single photon generation initially grows with $\Delta$, but tends back towards 1 as $\Delta \gg G$ as the decay into the subradiant bath dominates the system dynamics.

\hl{Finally, we study how superradiance within the multi-photon subspaces is impacted by the dynamical modulation. While it is in principle possible to recompute the modulation signal $\omega_c(t)$ by simulating an $N$ excitation problem where all the emitters are initially excited (demonstrated in the supplemental information), the cost of performing this simulation increases exponentially with $N$. However, physical intuition suggests that the pulse obtained by maximizing the single-photon superradiance facilitates the transfer of excitations between the disordered ensemble of emitters and the cavity mode, and hence should also enhance superradiance within the full $N-$excitation subspace. We see this effect in the photon emission fidelity in Fig.~\ref{fig:Fig4}a --- we point out that the enhancement obtained for the $N-$excitation problem is smaller than that obtained for the single-particle problem, since the pulse designed within the single-photon space is conceivably suboptimal for the $N-$excitation problem.}

\hl{An important question that arises for the $N-$excitation superradiance enhancement is whether it survives in the limit of large $N$. Since numerical simulations of the large $N$ model becomes prohibitively expensive, we instead consider the emitter frequencies to be chosen randomly only from a discrete set (Figs.~\ref{fig:Fig4}b and c) as opposed to a continuous distribution. Exploiting the permutational invariance of emitters at the same frequency, this system can be simulated with a cost that scales polynomially in the number of emitters but exponentially in the number of frequencies \cite{shammah2018open}. We expect the discrete probability distribution to at least qualitatively capture the properties of the continuous probability distribution, since it should in principle be possible discretize a continuous probability distribution over the emitter frequencies into bins whose widths depend on the linewidth of the system. While we do not have a rigorous proof of this statement, we numerically verify this for the problem of computing the fidelity metric in the supplement.  Figures \ref{fig:Fig4}b and c show the dependence of the fidelity on $N$ for $2$ and $4$ frequency problems, we find that the enhancement in superradiance on applying a pulse optimized in the single photon subspace remains nearly constant with $N$ in both cases, indicating that a multiplicative enhancement is possible even in  $N-$excitation superradiance.}

%overlap, Fig 4d. However, as the single photon approximation is less accurate with increasing $N$, the value of the optimized overlap likewise decreases.

In conclusion, we have studied a dynamically modulated Tavis-Cummings Hamiltonian with spectral disorder and presented an analysis of the emergence of superradiance in this model. Our conclusions indicate that superradiance can persist and be potentially technologically useful even in the limit of large spectral disorder, and that global quantum controls can be used to enhance it. These results are relevant to a number of on-going experimental efforts in studying and scaling-up solid-state quantum optical systems. One of the most important and interesting problems left open in our work is scaling quantum control designs, with experimentally realistic local or quasi-local controls, to multi-excitation subspaces of systems with large number of emitters. While we explored how controls designed with low excitation number subspaces can provide some improvement even in the high excitation number subspaces, there might be the potential to design controls within the low entanglement spaces of the Hilbert space (i.e.~states described by low bond dimension matrix product states), which have been investigated in the context of waveguide QED systems with spatial disorder \cite{manzoni2017simulating}.
\noindent %In this paper we introduced a quantum control technique to restore superradience to inhomogenously broadened emitter ensembles using a modulated cavity QED system. Through adjoint optimization of an applied drive, we are able to enhance the overlap of the output photon state of a collection of inhomogenous emitters with that of an ideal dicke superradiant state by several orders of magnitude. While a full optimization of a many emitter state is not yet possible without more clever classical simulations or access to a quantum computer, we provide a heuristic optimization that scales polynomially with the number of emitters and can still boost the superradiant overlap by two orders of magnitude.

%While we study the particular case of generating superradiant states, a similar approach could be fruitful for other problems which require the restoration of cooperative effects between inhomogeneous ensembles.

%Near term outlook about metrology?

%Discussion of future directions?

%Discussion of experimental feasibility?
% \\ \ \\

\bigskip

\noindent\textbf{Acknowledgments:} We thank Sattwik Deb Mishra and Daniel Malz for helpful discussions and Geun Ho Ahn and Eric Rosenthal for providing feedback on the manuscript. AW acknowledges the Herb and Jane Dwight Stanford Graduate Fellowship and the NTT Research Fellowship. RT acknowledges Max Planck Harvard research center for quantum optics (MPHQ) postdoctoral fellowship. KN acknowledges the Stanford Physics Undergraduate Research Program. We acknowledge support from the Department of Energy (DOE) grant number DE-SC0019174.\\

%\bibliography{Reference}
\bibliography{main.bbl}

% \clearpage

\renewcommand{\thefigure}{S\arabic{figure}}
\setcounter{figure}{0}

\pagebreak

\title{Supplemental Information}
\maketitle
\onecolumngrid

\noindent{\bf Numerical Techniques}

To simulate in the single-photon subspace, we use the time-dependent Schrodinger Equation:
\begin{flalign*}
    i\frac{d}{dt}\ket{\psi_{\text{cav}}(t)} & = (\omega_c(t) - i\frac{\kappa}{2})\ket{\psi_{\text{cav}}(t)} + g \sum_{i = 1}^N \ket{\psi_{\text{em}, i}(t)}\\
    i\frac{d}{dt}\ket{\psi_{\text{em}, i}(t)} & = (\omega_i - i\frac{\gamma}{2})\ket{\psi_{\text{em}, i}(t)} + g(N)\ket{\psi_{\text{cav}}(t)},
\end{flalign*}
where $\omega_c(t)$ is the time dependent frequency of the cavity, $\omega_i$ is the resonant frequency of the $i$th emitter, and $g(N)$ is the emitter-cavity coupling strength. The cavity emits into an output channel with decay rate $\kappa$, and the emitters in addition to coupling to the cavity also individually decay with decay rate $\gamma$.
We initialize the system to $\ket{\psi(0)} = \frac{1}{\sqrt{N}}\sum_{i = 1}^N \ket{\psi_{\text{em}, i}(t)}$.

To simulate in the multi-photon subspace, we use the time-dependent Master Equation:
\begin{flalign*}
    \frac{d}{dt}\rho(t) & = -i[H(t), \rho] + \sum_{j = 0}^N \left[L_j\rho L_j^\dagger - \frac{1}{2} \{L_j^\dagger L_j, \rho\} \right],\ \text{where} \\
    H(t) & = \omega_c(t) a^\dagger a + \sum_{i = 1}^{N}\omega_i\sigma_i^\dagger\sigma_i + g(N)\sum_{i = 1}^{N}(a^\dagger\sigma_i + \sigma_i^\dagger a),
    \\
    L_0 & = \sqrt{\kappa} a, \\
    L_j & = \sqrt{\gamma} \sigma_j \ \text{for } j = 1, 2 \dots N,
\end{flalign*}
where $\omega_c(t)$ is the time dependent frequency of the cavity, $\omega_i$ is the resonant frequency of the $i$th emitter, $a^\dagger$ is the raising operator of the cavity, $\sigma_i^\dagger$ is the excitation operator of the $i$th emitter, and $g(N)$ is the emitter-cavity coupling strength. Furthermore, we assume that the cavity emits into an output channel with decay rate $\kappa$, and the emitters in addition to coupling to the cavity also individually decay with decay rate $\gamma$.
We initialize the system to the excited state of all emitters.

\noindent{\bf Adjoint Optimization} 

We consider a general problem of a state $s(t)$ governed by 
\begin{equation}
    \frac{ds(t)}{dt} = A(t,p)s(t) + b(t),
\end{equation}
where $p$ are control parameters governing the trajectory of the system and $b(t)$ is the input to the system. The system at $t=0$ is at state $s(0)$ and the time window of interest is from $t=0$ to $t=T$. We consider an objective function $O$:
\begin{equation}
    O = f \left( \int_0^T \sigma^{\text{T}} (t) s(t) dt \right) ,
\end{equation}
where $f:\mathbb{C} \rightarrow \mathbb{R} $ is some function. We are interested in computing $\partial O/ \partial p_i$ efficiently.
 \\

\noindent\textit{Preliminaries:} Formally, the solution of the ODE can be written by introducing the propagator $U(t)$:
\begin{equation}
    U(t,p) = \mathrm{T}e^{\int_0^t A(\tau,p)d\tau}
\end{equation}
to reach
\begin{equation}
    s(t) = U(t,p)s(0) + U(t,p) \int_0^t U^{-1}(\tau, p)b(\tau)d\tau.
\end{equation}

An ODE for $U^{-1}(t,p)$ can also be calculated,
\begin{equation}
    \frac{d}{dt} U^{-1}(t,p) + U^{-1}(t,p)A(t,p) = 0.
\end{equation}

\noindent\textit{Calculation of gradient:} Consider perturbing $p_i \rightarrow p_i + \delta p_i$. This would result in the state $s(t) \rightarrow s(t) + \delta s(t)$, where
\begin{equation}
    \frac{d}{dt}\delta s(t) = U(t,p)\int_0^t U^{-1}(\tau, p) \frac{\partial A(\tau,p)}{\partial p_i} s(\tau) d\tau,
\end{equation}
the solution to which is given by
\begin{equation}
    \frac{\delta s(t)}{\delta p_i} = U(t,p)\int_0^t U^{-1}(\tau, p) \frac{\partial A(\tau,p}{\partial p_i}s(\tau)d\tau,
\end{equation}
where we have used the fact that $\delta s(0) = 0$. Now the objective function changes by 
\begin{equation}
    \frac{\delta O}{\delta p_i} = 2\mathrm{Re}\left\{ f'\left(\int_0^T \sigma^{\text{T}}(t) s(t) dt\right) \int_0^T \sigma^{\text{T}}(t) \frac{\delta s(t)}{\delta p_i} dt \right\}.
\end{equation}

Consider the term
\begin{flalign*}
    & \int_0^T \sigma^{\text{T}}(t) \frac{\delta s(t)}{\delta p_i} dt  \\
    & = \int_{t=0}^T\int_{\tau = 0}^t \sigma^{\text{T}}(t) U(t,p)U^{-1}(\tau, p)\frac{\partial A(\tau, p)}{\partial p_i} s(\tau) d\tau dt \\
    & = \int_{\tau=0}^T\int_{t = \tau}^T \sigma^{\text{T}}(t) U(t,p)U^{-1}(\tau, p)\frac{\partial A(\tau, p)}{\partial p_i} s(\tau) d\tau dt \\
    & = \int_{\tau = 0}^T a^{\text{T}}(t) \frac{\partial A(\tau, p)}{\partial p_i} s(\tau) d\tau,
\end{flalign*}
where we have defined
\begin{flalign*}
    a(\tau) = \int_{t = \tau}^T [U(t,p)U^{-1}(t,p)]^{\text{T}}\sigma(t)dt \\
    = U^{-\text{T}}(\tau, p) \int_{t = \tau}^T U^{\text{T}}(t,p) \sigma(t)dt.
\end{flalign*}

An ODE for $a(t)$ can now be derived:
\begin{align*}
\frac{d}{d\tau} a(\tau) &= \frac{d}{d\tau} U^{-\text{T}}(\tau, p) \int_{t = \tau}^T U^T(t,p) \sigma(t)dt  \\
& \qquad \qquad + U^{-\text{T}}(\tau,p)(-U^{\text{T}}(\tau,p)\sigma(\tau)) \\
&= [-U^{-1}(\tau,p)A(\tau,p)]^{\text{T}} \int_{t=\tau}^TU^{\text{T}}(t,p)\sigma(t)dt - \sigma(\tau)\\
&= -A^{\text{T}}(\tau, p) a(\tau) - \sigma(\tau).
\end{align*}
The boundary condition, as seen from the definition is simply $a(T) = 0$. This is the adjoint simulation. Finally, for the gradient we obtain
\begin{align}
    \frac{\delta O}{\delta p_i} =2\mathrm{Re}\left\{ f'\left(\int_0^T \sigma^{\text{T}}(t) s(t) dt\right) \int_0^T a^{\text{T}}(\tau) \frac{\partial A(\tau, p)}{\partial p_i} s(\tau) d\tau \right\}.
\end{align}
We note that the adjoint field is calculated only once, and all the derivatives $\delta O / \delta p_i$ are accessed immediately.

In our specific problem, the state evolution $s(t) = \psi(t)$ is governed by the system Hamiltonian, so
\begin{equation*}
    A(t,p) = H(t) = \omega_c(t) a^\dagger a + \sum_{i = 1}^{N}\omega_i\sigma_i^\dagger\sigma_i + g\sum_{i = 1}^{N}(a^\dagger\sigma_i + \sigma_i^\dagger a), 
\end{equation*}
and the free parameter $p$ corresponds to $\omega_c(t)$. Thus the gradient with respect to our one free parameter is
\begin{equation*}
    \frac{\partial A(t, p)}{\partial p} = a^\dagger a
\end{equation*}

We can calculate the output wave function by taking the overlap of the system wave function with the cavity state, and so our overlap objective is
\begin{equation*}
    \sigma^{\text{T}}(t) = f(t)\bra{\text{G}}a,
\end{equation*}
where $f(t)$ is the desired time dependent wavefunction (which we compute from the simulation of the homogeneous system).

As our objective function is the absolute value squared of the overlap, our gradients can be calculated as
\begin{equation*}
    \frac{\delta O}{\delta \omega_c(t)} =2\text{Re} \left\{ \bra{a(t)}a^\dagger a\ket{\psi(t)}  \int_0^T f^*(\tau)\braket{G|a|\psi(\tau)}^* d\tau  \right\}.
\end{equation*}

\noindent{\bf Large $N$ Model}

To derive a model for the limit of a large number of emitters, we first need to derive the system dynamics as a function of the emitter distribution.

We start with the single particle problem 
\begin{equation}
    \ket{\psi(t)} = \left[\alpha(t)a^\dagger + \sum_{\mu=1}^Ns_\mu(t)\sigma_\mu^\dagger \right]\ket{0}
\end{equation}
which is governed by
\begin{flalign*}
    i\dot{\alpha}(t) & = \left(\omega_c(t) - i \frac{\kappa}{2} \right) \alpha(t) + \frac{G}{\sqrt{N}}\sum_{\mu = 1}^N s_\mu(t) \\
    i\dot{s}_\mu(t) & = \delta_\mu s_\mu(t) + \frac{G}{\sqrt{N}}\alpha(t).
\end{flalign*}
We can integrate the second equation of motion 
\begin{equation*}
    s_\mu(t) = s_\mu(0)e^{-i\delta_\mu t} - i\frac{G}{\sqrt{N}} \int_0^t \alpha(s)e^{-i\delta_\mu(t-s)}ds.
\end{equation*}
Thus, 
\begin{equation*}
    \sum_{\mu = 1}^N s_\mu(t) = \sqrt{N}S_{in}(t) - i\sum_{\mu = 1}^N\frac{G}{\sqrt{N}}\int_0^t \alpha(s)e^{-i\delta_\mu(t-s)}ds,
\end{equation*}
where we have introduced
\begin{equation}
    S_{\text{in}}(t) = \sum_{\mu=1}^N S_\mu(0)\frac{e^{-i\delta_\mu t}}{\sqrt{N}}.
\end{equation}
Thus,
\begin{flalign*}
    i\dot{\alpha}(t) = \left(\omega_c(t) - i \frac{\kappa}{2}\right)\alpha &- i\frac{G^2}{N}\sum_{\mu = 1}^N \int_0^t \alpha(s)e^{-i\delta_\mu(t-s)}ds - iGS_{\text{in}}(t) \\
    i\dot{\alpha}(t) = \left(\omega_c(t) - i \frac{\kappa}{2}\right)\alpha &- iG^2 \int_0^t \alpha(s) K(t-s)ds - iGS_{\text{in}}(t),
\end{flalign*}
where for $t \geq 0$
\begin{equation*}
    K(t) = \frac{1}{N}\sum_{\mu = 1}^N e^{-i\delta_\mu t}.
\end{equation*}

We now have an expression for the cavity dynamics as a function of the emitter frequencies. Until this point, the system of equations is exact. Assuming each frequency is picked independently at random, in the limit of large $N$, $K(t)$ can be expressed as
\begin{equation}
    K(t) = \braket{e^{-i\delta t}} = \int P(\delta)e^{-i\delta t} d\delta.
\end{equation}

A relatively simple model can be derived if $P(\delta$) is Lorentzian,
\begin{equation*}
    P(\delta) = \frac{\delta_0}{\pi}\frac{1}{\delta^2 + \Delta_0^2}.
\end{equation*}
In this case, for $t \geq 0$
\begin{flalign*}
    K(t) &= \frac{\delta_0}{\pi} \int_{-\infty}^\infty \frac{e^{-i\delta t}}{\delta^2 + \Delta_0^2}d\delta \\
    &= \frac{1}{2\pi i} \int_{-\infty}^\infty \left( \frac{1}{\delta-i\Delta_0} - \frac{1}{\delta+i\Delta_0} \right)e^{-i\delta t}d\delta \\
    &= e^{-\Delta_0 t}.
\end{flalign*}

Now, introducing $\beta(t)$ as
\begin{flalign*}
    \beta(t) &= G\int_0^t \alpha(s) K(t-s) ds \\
    &= G\int_0^t e^{-\Delta_0(t-s)}\alpha(s)ds,
\end{flalign*}
we can rewrite the equations of motion as 
\begin{flalign*}
    \dot{\alpha}(t) &= -i\left(\omega_c(t) - i \frac{\kappa}{2}\right)\alpha - G\beta(t) - iGS_{\text{in}}(t) \\
    \dot{\beta}(t) &= -\Delta_0\beta(t) + G\alpha(t)
\end{flalign*}
which is a model of a coupled cavity (Fig 3a). Here, $\alpha(t)$ still represents the original cavity, and the new $\beta(t)$ represents the superradiant state. The decay rate $\Delta_0$ of the superradiant state $\beta(t)$ can be viewed as the decay into a bath consisting of the subradiant states, whose coupling rate to the cavity are 0 in the limit of large $N$.

To simulate an initial state which is an equal superposition of all excited states, we would use $S_\mu(0) = 1/ \sqrt{N}$ with which
\begin{flalign*}
    S_{\text{in}}(t) &= \frac{1}{N}\sum_\mu e^{-i\delta_\mu t} \\
    &= \braket{e^{-i\delta t}} \\
    &= \int P(\delta)e^{-i \delta t}d\delta \\
    &= e^{-\Delta_0 t}u(t),
\end{flalign*}
where $u(t)$ is the unit step function. As $S_{in}(t)$ is the only input to the system, it bounds the maximum single photon generation fidelity. As the total power entering into the system, $\int_0^\infty GS_{in}(t)dt$, scales as $G/\Delta_0$, the maximum possible fidelity should scale accordingly.
 \\

\vspace{10pt}

\noindent{\bf Multi-photon Optimization}

By maximizing the overlap of the full $N$-photon output wavepacket with that of the homogeneous case, we can enhance the emission rate of 2 to 5 emitter inhomogeneous ensembles, Sup. Fig. 1. However, this optimization procedure is far too expensive for larger $N$, as the simulation costs scales exponentially.

\begin{figure}[h!]
\centering
\includegraphics[width=0.6\linewidth]{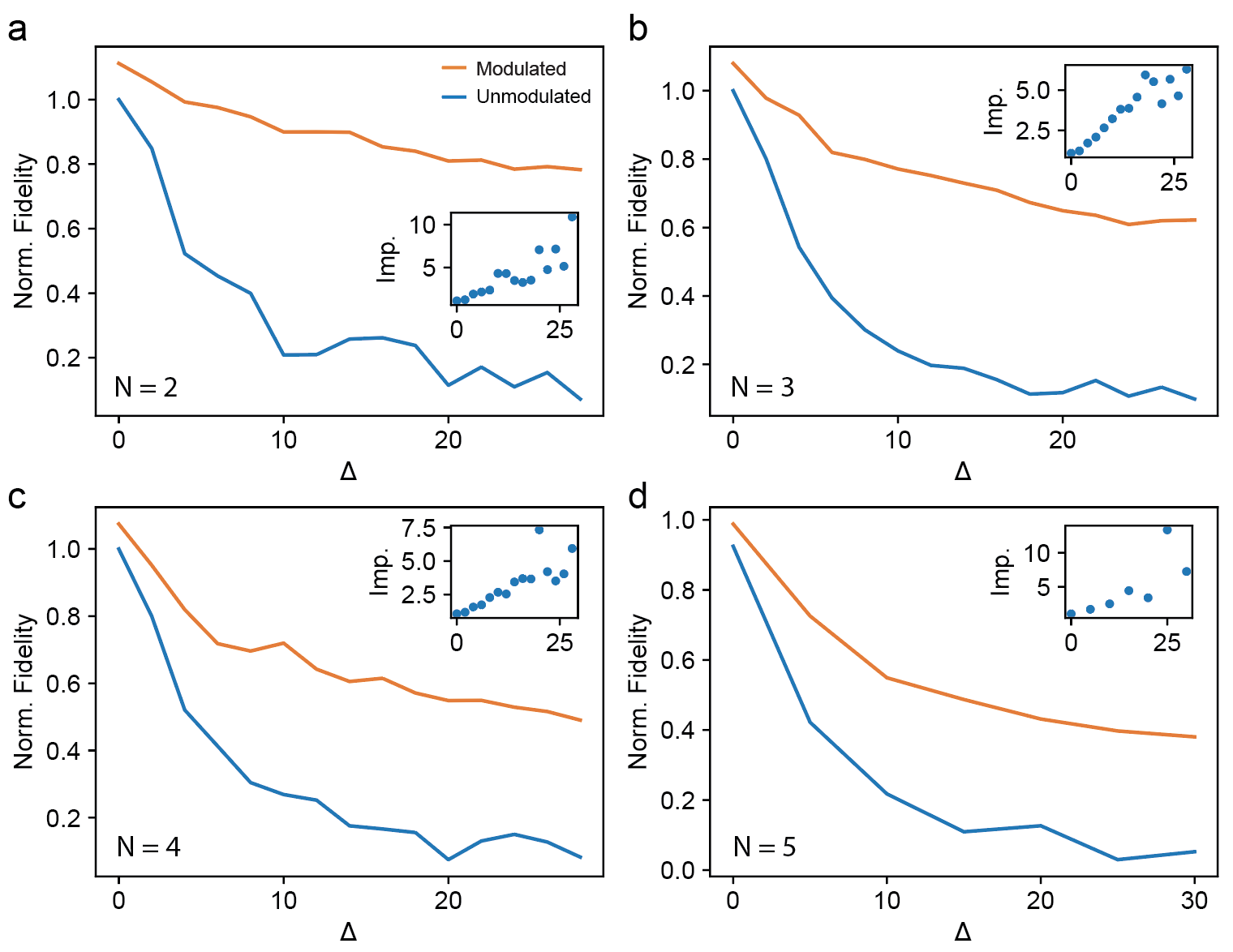}
%\captionsetup{singlelinecheck=no, justification = RaggedRight}
\caption{{\bf{$N$ Photon Multi-Emitter Optimization}} \textbf{(a-d)} $N$ photon space optimization of $N = 2,3,4,5$ emitters with a nonradiative decay rate of $0.5\kappa$ and $\Delta$ in units of $\kappa$. All data points are an average of 25 ensembles.}
\end{figure}

To simulate larger ensembles more efficiently, we can bin the emitter frequencies in order to maximize permutational invariance, allowing the simulation to scale polynomially with the number of emitters and exponentially only with the number of frequency bins. To validate that this discretization can still capture the desired system dynamics, we calculate the simulation error in single photon generation fidelity with increasing $N$, Sup. Fig. 2a. Here we define the simulation error as the fractional difference in fidelity between the discretized ensemble and full ensemble. We find that the error induced through discretization can be made to be small with enough bins, and also that the number of bins needed for a given error threshold does not increase with the number of emitters (given that g scales as $1/\sqrt{N}$). For a given $N$, we also find that the error introduced by binning decreases as $g$ is increased, Sup. Fig. 2b,likely because the linewidths of the emitters grow, increasing the minimum linewidth of the system.

\begin{figure}[h!]
\centering
\includegraphics[width=0.65\linewidth]{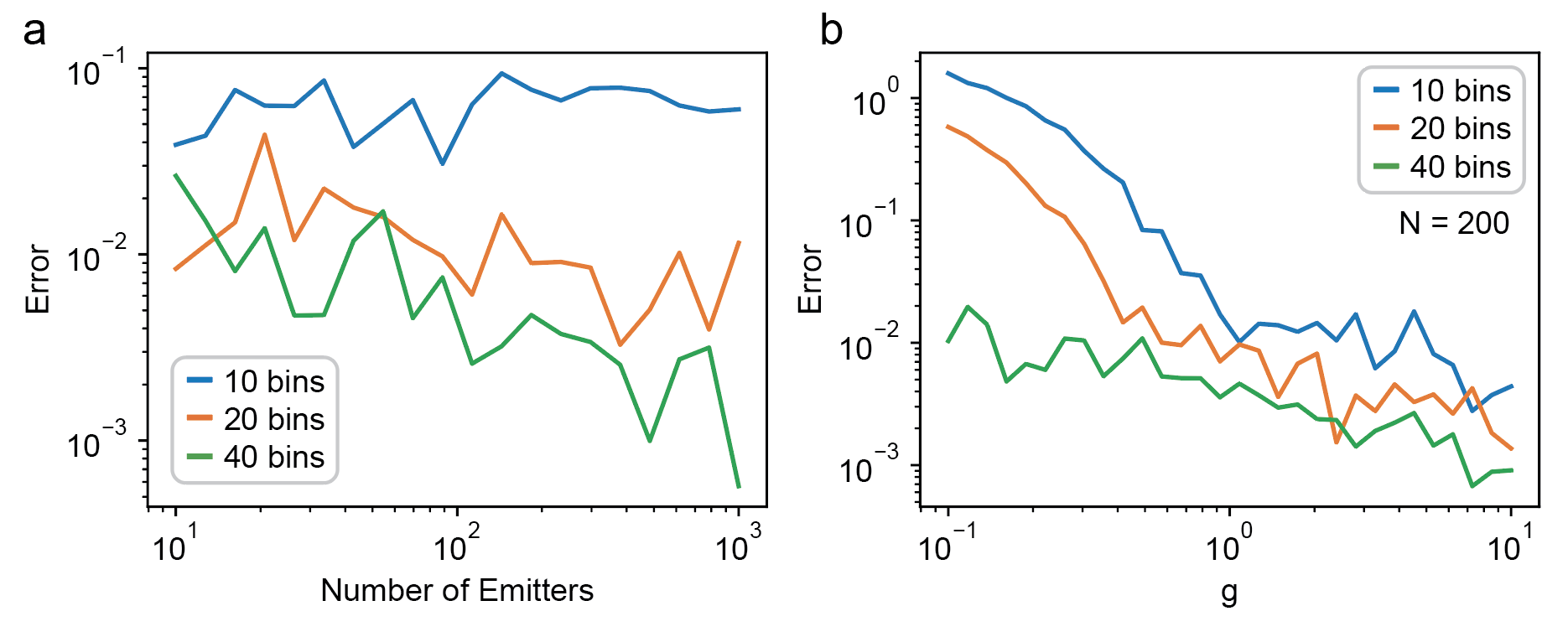}
%\captionsetup{singlelinecheck=no, justification = RaggedRight}
\caption{{\bf{Emitter Binning}} \textbf{(a)} Error in calculated single photon generation fidelity due to binning as N increases. Here $\Delta = 10\kappa$, $g = \kappa/\sqrt{N}$. \textbf{(b)} Error in calculated single photon generation fidelity due to binning as $g$ increases for $N = 200$. $g$ is in units of $\kappa/\sqrt{200}$.}
\end{figure}

\vspace{1in}

\end{document}